\documentclass[a4paper,12pt]{article}
\pdfoutput=1 

\usepackage{jheppub} 

\usepackage[T1]{fontenc} 
\usepackage{orcidlink}
\usepackage{lipsum}
\usepackage{float}
\usepackage{hyperref}
\usepackage{tcolorbox}
\usepackage{amssymb}
\usepackage{multicol}
\usepackage{graphicx}
\usepackage{color}
\usepackage{caption}
\usepackage{subcaption}
\usepackage{flexisym}
\usepackage{listings}
\usepackage{cancel}
\usepackage{colortbl}
\usepackage{graphicx}
\usepackage{epstopdf}

\usepackage{amsmath}
\usepackage{jheppub}

\usepackage{mathtools}

\title{\boldmath Search for $B{}^0_s \rightarrow \ell^{\mp} \tau^{\pm}$ with the Semi-leptonic Tagging Method at Belle}

\preprint{\vbox{ \hbox{   }
					    	\hbox{Belle Preprint 2023-02 }
                        	\hbox{KEK Preprint 2022-49} 
                     }}

\collaboration{The Belle Collaboration}
  \author{L.~Nayak\,\orcidlink{0000-0002-7739-914X},} 
  \author{S.~Nishida\,\orcidlink{0000-0001-6373-2346},} 
  \author{A.~Giri\,\orcidlink{0000-0002-8895-0128},} 
  \author{I.~Adachi\,\orcidlink{0000-0003-2287-0173},} 
  \author{H.~Aihara\,\orcidlink{0000-0002-1907-5964},} 
  \author{D.~M.~Asner\,\orcidlink{0000-0002-1586-5790},} 
  \author{H.~Atmacan\,\orcidlink{0000-0003-2435-501X},} 
  \author{V.~Aulchenko\,\orcidlink{0000-0002-5394-4406},} 
  \author{T.~Aushev\,\orcidlink{0000-0002-6347-7055},} 
  \author{R.~Ayad\,\orcidlink{0000-0003-3466-9290},} 
% ...

\newcommand{\fbi}{\mathrm{fb}^{-1}}
\newcommand{\Bs}{B_s{}}

\newcommand{\Bszero}{B^0_s}
\newcommand{\Bszerobar}{\overline{B}{}^0_s}
\newcommand{\DE}{\Delta E}
\newcommand{\EB}{{E_B^*}}
\newcommand{\Ebeam}{E^*_\mathrm{beam}}
\newcommand{\effsig}{\epsilon_{\mathrm{sig}}}
\newcommand{\GeV}{\mathrm{GeV}}
\newcommand{\MeV}{\mathrm{MeV}}
\newcommand{\KS}{K_S^0}
\newcommand{\Mbc}{M_\mathrm{bc}}
\newcommand{\Nbkg}{N_{\mathrm{bkg}}^{\mathrm{exp}}}
\newcommand{\Nobs}{N_{\mathrm{obs}}}
\newcommand{\NBs}{N_{B_s}}
\newcommand{\OFBDT}{\mathcal{O}_\mathrm{FastBDT}}
\newcommand{\pb}{\mathrm{pb}}

\newcommand{\piz}{\pi^0}
\newcommand{\pOne}{p^*_1}
\newcommand{\pTwo}{p^*_2}
\newcommand{\pThree}{p^*_3}

\abstract{We present a search for the lepton-flavor-violating decays $B{}^0_s \rightarrow \ell^{\mp}\tau^{\pm}$, where $\ell = e, \mu$, using the full
data sample of $121~\mathrm{fb}^{-1}$ collected at the $\Upsilon(5S)$ resonance with the Belle detector at the KEKB asymmetric-energy $e^+e^-$ collider. We use $B{}^0_s \overline{B}{}^0_s$ events in which one $B{}^0_s$ meson is reconstructed in a semileptonic decay mode and the other in the signal mode. We find no evidence for $B{}^0_s \rightarrow \ell^{\mp}\tau^{\pm}$ decays and set upper limits on their branching fractions
at $90\%$ confidence level as $\mathcal{B}(B{}^0_s \rightarrow e^{\mp}\tau^{\pm}) < 14 \times 10^{-4}$ and $\mathcal{B}(B{}^0_s \rightarrow \mu^{\mp}\tau^{\pm}) < 7.3 \times 10^{-4}$. Our result represents the first upper limit on the $B{}^0_s \rightarrow e^{\mp}\tau^{\pm}$ decay rate.}


\begin{document} 
\maketitle
\flushbottom

\section{Introduction}

The lepton-flavor-violating (LFV) decays $B{}^0_s \rightarrow \ell^{\mp} \tau^{\pm}$,  
where $\ell= e, \mu$, are forbidden  
in the standard model (SM). Such decays can occur via neutrino mixing by loop and box diagrams~\cite{Super-Kamiokande:2000ywb}, but the predicted decay rates are far below current experimental capabilities. Thus, any observations at current experiments would constitute an unambiguous signature of new physics (NP).
Recent results indicating possible lepton flavor universality violation in $B$ meson decay have
been discussed in Refs.~\cite{London:2021lfn,Egede:2022rxc}, where many NP models are proposed to explain it.
Such models allow significantly enhanced LFV decay rates that may be detectable with current facilities.
For example, the models containing a heavy neutral gauge boson ($Z'$) could lead to an enhanced $B{}^0_s \rightarrow \mu^-\tau^+$ branching fraction, up to $10^{-8}$ when only left- or right-handed couplings to quarks are considered, or of order $10^{-6}$ \cite{Crivellin:2015era}, if both are allowed. 
In models with either scalar or vector leptoquarks,
the prediction for the branching fraction of
$B{}^0_s \rightarrow \ell^-\tau^+$ can be as large as $10^{-5}$~\cite{Smirnov:2018ske,deMedeirosVarzielas:2015yxm,Becirevic:2016oho}, depending on the assumed leptoquark mass. 
It is imperative to search for signals of physics beyond the SM in all possible avenues, and
since the expected branching fraction of $B{}^0_s \rightarrow e^-\tau^+$ may differ from $B{}^0_s \rightarrow \mu^-\tau^+$ depending on models, it is important to search for both decay modes to obtain additional information regarding the NP. To date, no experimental results for $B{}^0_s \to e^{\mp}\tau^{\pm}$ have been reported while an upper limit $\mathcal{B} (B{}^0_s \rightarrow \mu^{\mp} \tau^{\pm}) < 3.4 \times 10^{-5}$ at $90\%$ confidence level (CL)~\cite{LHCb:2019ujz} has been reported by LHCb.

In this paper, we report a search for $B{}^0_s \rightarrow \ell^{\mp} \tau^{\pm}$ decays using $121~\fbi$ of data  collected by the Belle experiment at the KEKB asymmetric-energy $e^+e^-$ collider~\mbox{\cite{Kurokawa:2001nw,Abe:2013kxa}}. The data were collected at an $e^+e^-$ center-of-mass (c.m.)\  energy corresponding to the $\Upsilon$(5S) resonance mass. 
\section{Data sample and Belle detector}

The Belle detector 
is a large-solid-angle magnetic spectrometer comprising a silicon vertex detector, a 50-layer central drift chamber (CDC), an array of aerogel threshold Cherenkov counters (ACC), a barrel-like arrangement of time-of-flight
scintillation counters (TOF), and a CsI(Tl) crystal electromagnetic calorimeter (ECL). All these components are located inside a superconducting solenoid providing a magnetic field of 1.5 T. An iron flux return located outside
the solenoid coil is instrumented with resistive plate chambers to detect $K^0_L$ mesons and muons (KLM). A more detailed description of 
the detector and its layout and performance can be found in
Refs.~\mbox{\cite{Belle:2000cnh,10.1093/ptep/pts072}}.

We study the properties of signal events, identify sources of background, and optimize selection criteria using Monte Carlo (MC) simulated events. These samples are generated using 
EvtGen~\cite{Lange:2001uf}. The detector response is simulated using the Geant3 framework~\cite{Brun:1987ma}. We simulate 20 million $\Bszero \rightarrow \ell^{\mp} \tau^{\pm}$ MC events to study the detector response and to calculate signal reconstruction efficiencies.
To estimate backgrounds, we use MC samples of 
$B^{(\ast)}_s\overline{B}{}^{(\ast)}_s$ and $B^{(\ast)}_{u,d}\overline{B}{}^{(\ast)}_{u,d}X$ events, with $B_s \leftrightarrow \overline{B}_{s}$,
$B_{d} \leftrightarrow \overline{B}_{d}$ mixing~\cite{Workman:2022ynf}, and $e^+ e^- \rightarrow q\overline{q}$ $(q=u,d,s,c)$ events.  
These samples, referred to as generic MC, are equivalent to six times the data luminosity. The Belle data are converted into the Belle II format~\cite{Gelb:2018agf}, and the particle and event reconstruction are performed within the basf2 framework~\cite{Kuhr:2018lps,the_belle_ii_collaboration_2021_5574116} of the Belle II experiment.

The $\Bszero$ and $\Bszerobar$ mesons are produced in the process
$e^+e^- \to \Upsilon(5S) \to B{}^{(\ast)0}_s\overline{B}{}^{(\ast)0}_s$,
with $B^{*0}_s \to B{}^0_s\gamma$, $\overline{B}{}^{*0}_s \to \overline{B}{}^0_s\gamma$ and with fast $B^{0}_s \leftrightarrow \overline{B}{}^{0}_s$ mixing, such that half of the events contain same flavor $\Bs$ pairs. 
The $\Upsilon(5S)$ resonance production cross section is $340 \pm 16~\pb$~\cite{Belle:2012tsw},
and $f_s$, its total branching fraction for decays to $B^{(\ast)0}_s\overline{B}{}^{(\ast)0}_s$,
is $0.201 \pm 0.031$~\cite{Workman:2022ynf}. Therefore, the Belle data sample 
is estimated to contain $(16.6 \pm 2.7) \times 10^6$ $\Bs$ mesons.
\section{Event Selection and Analysis Overview}
Hereafter, $B_s$ refers to either $\Bszero$ or $\Bszerobar$,
and the inclusion of charge-conjugated modes is implied.
In this analysis, one $B_s$ is reconstructed in a semileptonic decay  mode $\Bszerobar \to D_s^+\ell^-(X) {\overline{\nu}_{\ell}}$ and used as a tag,
where $X$ stands
for any particles such as $\pi$ or 
a combination of pions, and the signal $B_s \to \ell^-\tau^+$ is searched for in the mode $\tau^+ \to \ell^+\overline{\nu}_\tau\nu_\ell$.
We label the primary and secondary leptons from the $\tau$ decay on the signal side 
$B_s$ as $\ell_1$ and $\ell_2$, and the lepton
on the tag side as $\ell_3$.
In short, we search for $B_s \to \ell_1^-\tau^+$
($\to \ell_2^+ \overline{\nu}_\tau \nu_{\ell_2}$) 
with $\Bszerobar$ tagged by the decay
$\Bszerobar \to D_s^+\ell_3^-(X){\overline{\nu}_{\ell_3}}$.
\begin{figure}[tbh]
\centering
\includegraphics[width= 0.8\linewidth]{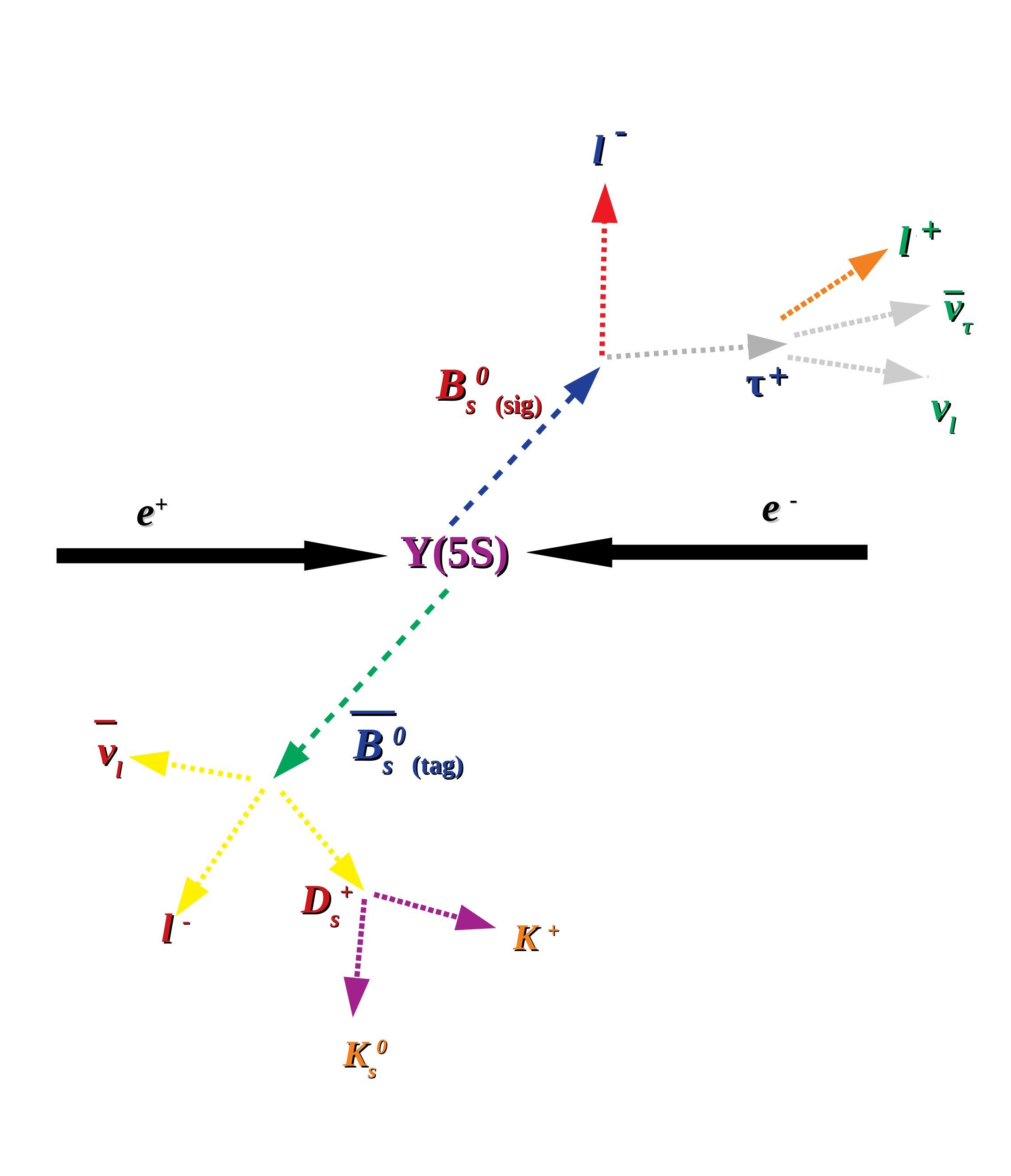}

\caption{{A schematic diagram of the process under study}, separated into signal and tag sides.}
\label{fig: diagram}
\end{figure}
Figure~\ref{fig: diagram} shows a schematic diagram of the 
process, separated into signal and tag sides. To avoid biasing the results, all selection criteria are determined in a ``blind'' manner, i.e., they are optimized using MC samples only, before 
the experimental data in the signal region are revealed.

For charged particles, aside  
for pions 
from $\KS$, the distance of nearest approach of the track perpendicular to
and along the beam direction, with respect to the nominal interaction point, are required to be
less than 0.5 cm and 2.0 cm, respectively. The $\KS$ candidates are reconstructed by combining two oppositely charged particles (assumed to be pions) with an invariant mass between 487 and $508~\MeV/c^2$; this range corresponds to
approximately three standard deviations ($\pm 3\sigma$) in the invariant mass
resolution around the nominal $\KS$ mass~\cite{Workman:2022ynf}. Such candidates are further subjected to a neural network-based identification
~\cite{Belle:2018xst}. 
The $\piz$ candidates are reconstructed from pairs of photons detected as ECL clusters without any associated charged tracks in the
CDC.  
The energy of each photon is required to be greater than $50~\MeV$ if the photon is detected in the barrel region ($32.2^\circ < \theta < 128.7^\circ$, where $\theta$ is its polar angle), greater than $100~\MeV$ if the photon is in the forward endcap region ($12.4^\circ < \theta < 31.4^\circ$), and greater than $150~\MeV$ if the photon is in the backward endcap region ($130.7^\circ < \theta < 155.1^\circ$)~\cite{Belle:2000cnh,10.1093/ptep/pts072}. 
The invariant mass of each photon pair is required to be between 120 and $150~\MeV/c^2$; this range corresponds to a window of approximately $\pm 3\sigma$ in the invariant mass resolution 
around the nominal $\piz$ mass, and the reconstructed $\piz$ momentum in the c.m.\ frame ($p^*_\piz$) must be 
greater than $0.2~\GeV/c$. A mass constrained fit to the nominal $\piz$ mass is performed to improve momentum resolution.

To identify charged hardons, we use information on the light yield from the ACC, 
crossing time from the TOF, and specific ionization from the CDC. This information is combined into likelihoods $\mathcal{L}_K$ and $\mathcal{L}_\pi$ for a given track to be a $K^+$ or $\pi^+$, respectively. To identify $K^+$ or $\pi^+$ tracks, we require $\mathcal{L}_{K}/(\mathcal{L}_K+\mathcal{L}_\pi) > 0.6$ or $\mathcal{L}_{\pi}/(\mathcal{L}_K+\mathcal{L}_\pi) > 0.6$. This requirement is more than 93$\%$ efficient in identifying pions, with a 
$K^+$ mis-identification rate below $5\%$. Muon candidates are selected based on information from the KLM~\cite{Abashian:2002bd}. We calculate a normalized muon likelihood ratio $\mathcal{R}_\mu = \mathcal{L}_{\mu}/(\mathcal{L}_{\mu}+\mathcal{L}_{\pi}+\mathcal{L}_K)$, where $\mathcal{L}_{\mu}$ is the likelihood for muons, and require $\mathcal{R}_\mu > 0.9$. This requirement has an efficiency of $85-92\%$ and a probability of misidentifying a hadron as a muon below $7\%$. Electron candidates are identified using the ratio of calorimetric cluster energy to particle momentum, the shower shape in the ECL, the matching of the track with the ECL cluster, the specific ionization in the CDC, and the number of photoelectrons in the ACC~\cite{Hanagaki:2001fz}. This information is used to calculate a normalized electron likelihood ratio $\mathcal{R}_e = \mathcal{L}_{e}/(\mathcal{L}_{e}+\mathcal{L}_{\text{had}})$, where $\mathcal{L}_{e}$ is the likelihood for electrons and $\mathcal{L}_{\text{had}}$ is a product of hadron likelihoods. We require $\mathcal{R}_e > 0.9$. 
This requirement has an efficiency of $84 - 92\%$ and a probability of misidentifying a hadron as an electron below 1$\%$.

For the signal side $B_s \to \ell_1^-\tau^+$ ($\to \ell_2^+ \overline{\nu}_\tau \nu_{\ell_2}$),
we require that the two leptons $\ell_1$ and $\ell_2$ have opposite charges and that $\pOne > \pTwo$, $\pOne > \pThree$, and $\pOne > 1.9~\GeV/c$, 
where $\pOne$, $\pTwo$ and $\pThree$ are the momenta of $\ell_1$,
$\ell_2$ and $\ell_3$ in the c.m.\ frame.
To suppress background coming from $J/\psi \rightarrow \ell^+\ell^-$, the candidate is rejected if the invariant mass of the two leptons 
$M_{\ell_1\ell_2}$ satisfies 
$M_{\ell_1\ell_2} \in [3.01,3.12]~\GeV/c^2$
for the $B_s \to e^-\tau^+ (\to e^+\overline{\nu}_\tau\nu_e)$ mode,
and 
$M_{\ell_1\ell_2} \in [3.05,3.12]~\GeV/c^2$
for the $B_s \to \mu^-\tau^+ (\to \mu^+\overline{\nu}_\tau\nu_\mu)$ mode.  
The wider asymmetric veto interval for the electron mode is due to bremsstrahlung energy loss.

\begin{figure}[tbh]
 \begin{center}
 \includegraphics[width=77mm]{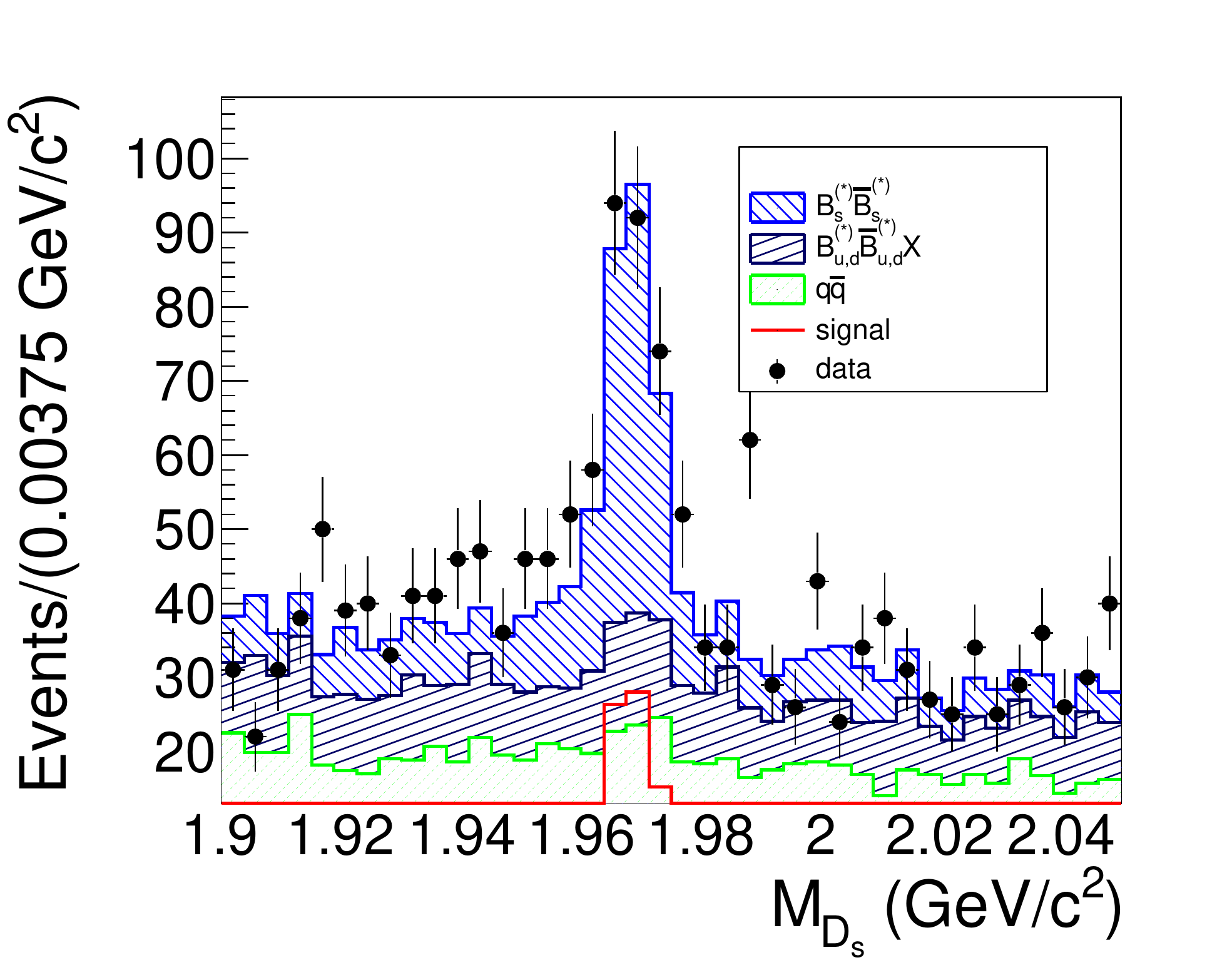}
 \end{center}
 \caption{The $M_{D_s}$ distribution of signal MC, generic MC and data. The different 
 background component in generic MC are indicated by different colours as shown in the legend. The MC samples are normalized with respect to the data luminosity. The signal component corresponds to $\mathcal{B} = 1 \times 10^{-2}$.}
\label{fig: dsmass}
\end{figure}

\begin{figure}[tbh]
\centering
\begin{multicols}{2}
\begin{subfigure}[t]{\linewidth}{}
\centering
\includegraphics[width= \linewidth]{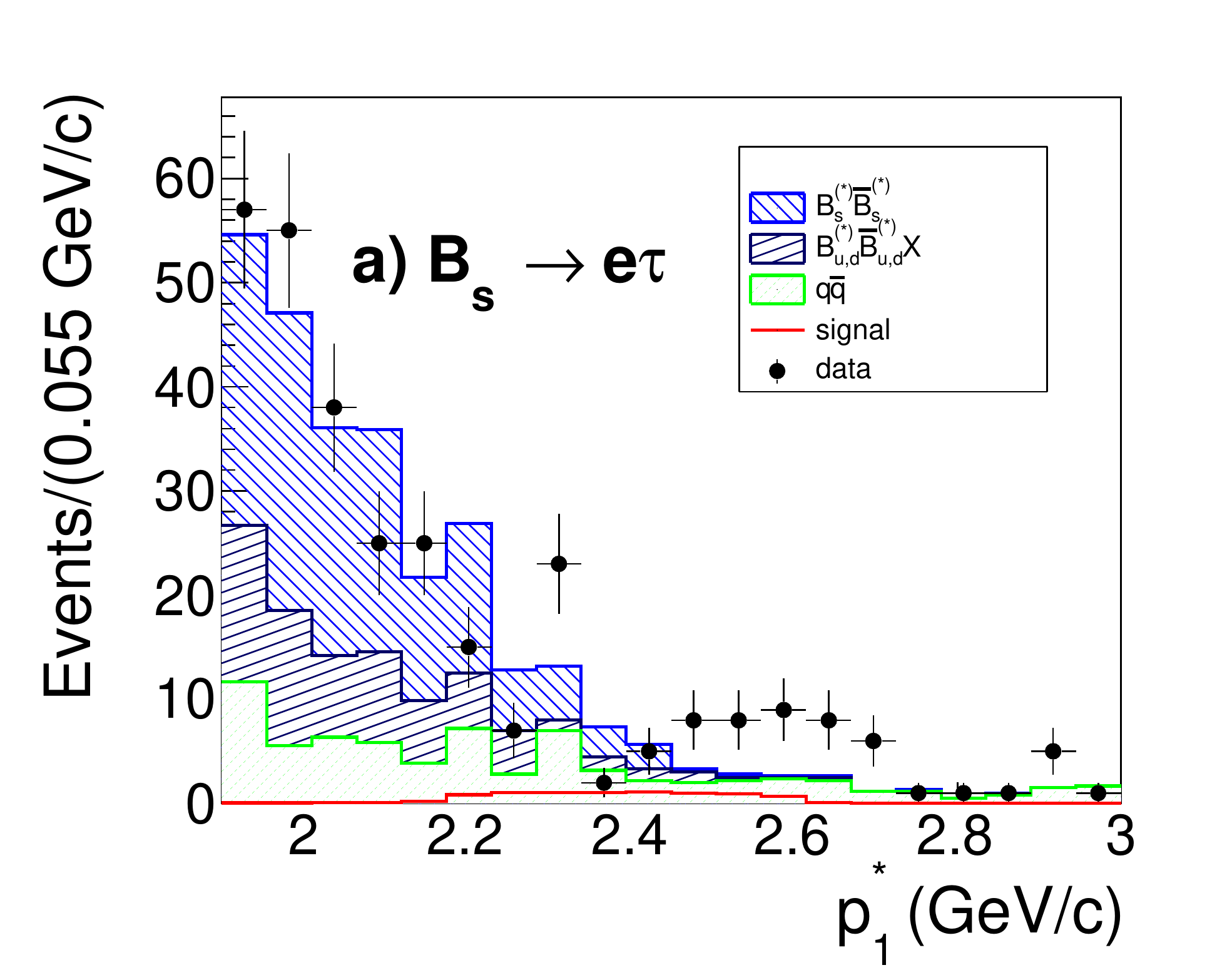}
\end{subfigure}

\begin{subfigure}[t]{\linewidth}{}
\centering
\includegraphics[width= \linewidth]{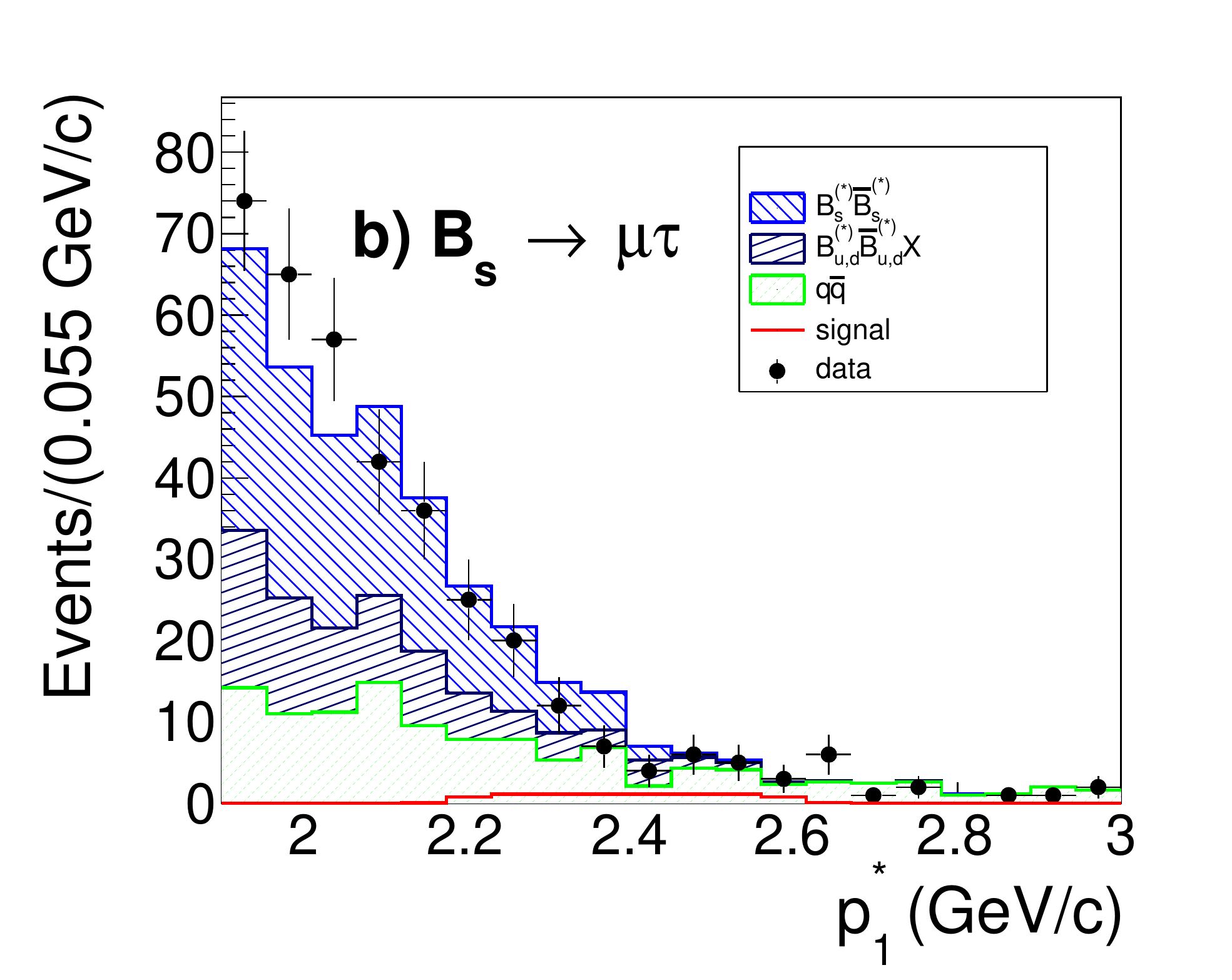}
\end{subfigure}

\end{multicols}
\caption{The $p_{1}^*$ distribution of signal MC, generic MC and data in $B_s \rightarrow e^-\tau^+$ (a) and $B_s \rightarrow \mu^-\tau^+$ (b) modes.  The different 
 background components in generic MC are indicated by different colours as shown in the legend. The MC samples are normalized with respect to the data luminosity. The signal components correspond to $\mathcal{B} = 1 \times 10^{-3}$.}
\label{fig:p1-nomva}

\end{figure}

For the tag side $\overline{B}{}^0_s \to D_s^+\ell_3^-(X)\overline{\nu}_{\ell_3}$, the charge of $\ell_3$ can be opposite to or the same as $\ell_1$, as $B_s$ mixing produces equal numbers of opposite and same charge combinations. However, we accept only combinations where the charges of $\ell_1$ and $\ell_3$ are the same; this significantly reduces combinatorial background. We reconstruct $D_s$ meson candidates with opposite charge to $\ell_3$ from the following five decay modes: $D_s^{+} \rightarrow \phi\pi^+$, $\overline{K}{}^{*0}K^+, \phi\rho^0\pi^+, \KS K^+$
and $\phi\rho^+$.
Here, $\rho^0$, $\rho^+$, $\overline{K}{}^{*0}$ and $\phi$ are reconstructed
through $\rho^0 \to \pi^+\pi^-$, {{$\rho^+ \to \pi^+\pi^0$}}, $\overline{K}{}^{*0} \to K^-\pi^+$ and $\phi \to K^+K^-$,
and candidates are required to have a reconstructed invariant mass 
$625~\MeV/c^2 < M_{\pi^+\pi^{-}(\pi^+\pi^0)} < 925~\MeV/c^2$ for $\rho^0$($\rho^+$), $845~\MeV/c^2 < M_{K^+\pi^{-}} < 945~\MeV/c^2$ for $\overline{K}{}^{*0}$, and 
$1.01~\GeV/c^2 < M_{K^+K^{-}} < 1.03~\GeV/c^2$ for $\phi$.
The $D_s$ candidate is then combined with an
$e$ or $\mu$ to form a $B_s$ meson 
candidate. Figure~\ref{fig: dsmass} shows the mass distribution of $D_s^+$ meson candidates. 
The mass of the $D_s^+$ candidate is required to be between 1.96 and $1.98~\GeV/c^2$. These mass windows correspond to $\pm 3\sigma$ in the invariant mass resolution around the nominal masses~\cite{Workman:2022ynf}. Figure~\ref{fig:p1-nomva} shows the $\pOne$ distribution for $B_s \to e^-\tau^+$
and $B_s \to \mu^-\tau^+$ after initial selections.

The background comes from the continuum $e^+e^- \rightarrow q\overline{q}$ 
process and $e^+e^- \to B{}^{(\ast)0}_s\overline{B}{}^{(\ast)0}_s, B{}^{(\ast)}\overline{B}{}^{(\ast)}X$. 
The continuum events have final-state particles momenta spatially correlated
in two directions forming jet-like structures, while 
particles from $B{}^{(\ast)0}_s\overline{B}{}^{(\ast)0}_s$ events
are distributed almost uniformly over the full 
solid angle in the c.m.\ frame.
We use this difference in the topology to suppress the continuum background.
The background from 
$B{}^{(\ast)0}_s\overline{B}{}^{(\ast)0}_s$ or $B{}^{(\ast)}\overline{B}{}^{(\ast)}X$ 
are suppressed using other variables characterizing the signal
decay chains. We form a single FastBDT~\cite{Keck:2017gsv} classifier trained using simulated samples with the following discriminating variables as input:
$p_2^*$; $p_3^*$; the extra energy from the tracks and clusters not
used for signal and tag reconstruction in the calorimeter;
the sum of the energy of the clusters and charged tracks in the c.m.\ frame;  
the missing energy which is the absolute value of the sum of the four-momenta of all charged candidates; the invariant mass squared of the sum of the four-momenta of all charged candidates; 
%
$(2 \Ebeam E^*_{D_s \ell_3}- m^2_{B_s}c^4 - M^2_{D_s \ell_3}c^4)/(2 |\vec{p}^{\,*}_{B_s}| |\vec{p}^{\,*}_{D_s\ell_3}| c^2)$;
the cosine of the angle between $p_{1}^*$ and $p_{2}^*$;
the mass of the $D_s^+$ candidate; and 16 modified Fox-Wolfram moments~\cite{Fox:1978vu,PhysRevLett.91.261801} calculated from the signal $B_s$ daughters and the particles from the rest of the event. 
Here, $\Ebeam$ is the beam energy in the c.m.\ frame, 
$m_{B_s}$ is the nominal $B_s$ mass,
$|\vec{p}^{\,*}_{B_s}| = \left(\sqrt{E^{*2}_{\mathrm{beam}} - m^2_{B_s}c^2}\right)/c $,
and
$E^*_{D_s \ell_3}$, $\vec{p}^{\,*}_{D_s\ell_3}$
and $M_{D_s\ell_3}$ are calculated from the reconstructed 
$D_s\ell_3$ system.

\begin{figure}[htb]
\centering
\begin{multicols}{2}
\begin{subfigure}[t]{\linewidth}{}
\centering
\includegraphics[width= 1.09\linewidth]{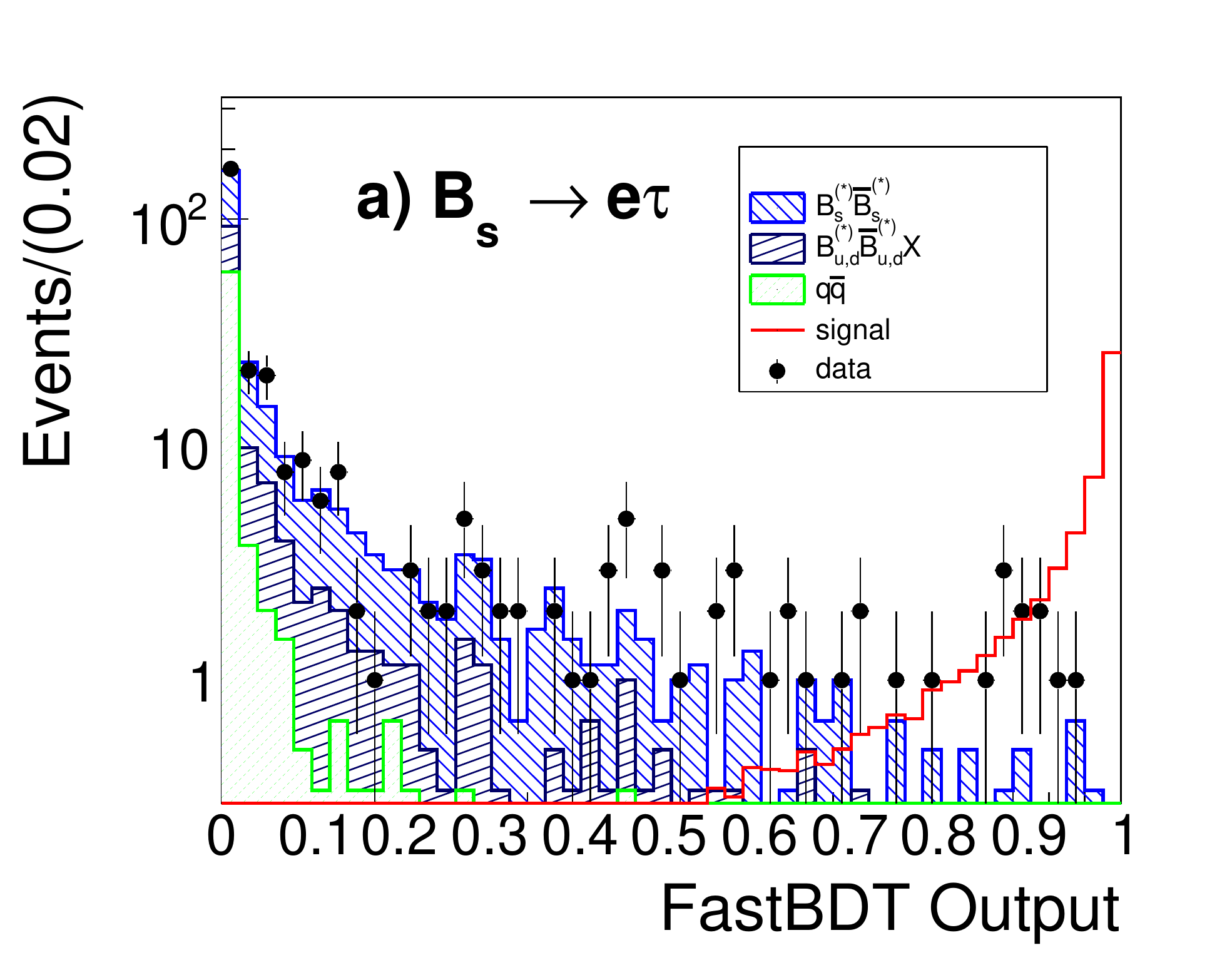}

\end{subfigure}

\begin{subfigure}[t]{\linewidth}{}
\centering
\includegraphics[width= 1.09\linewidth]{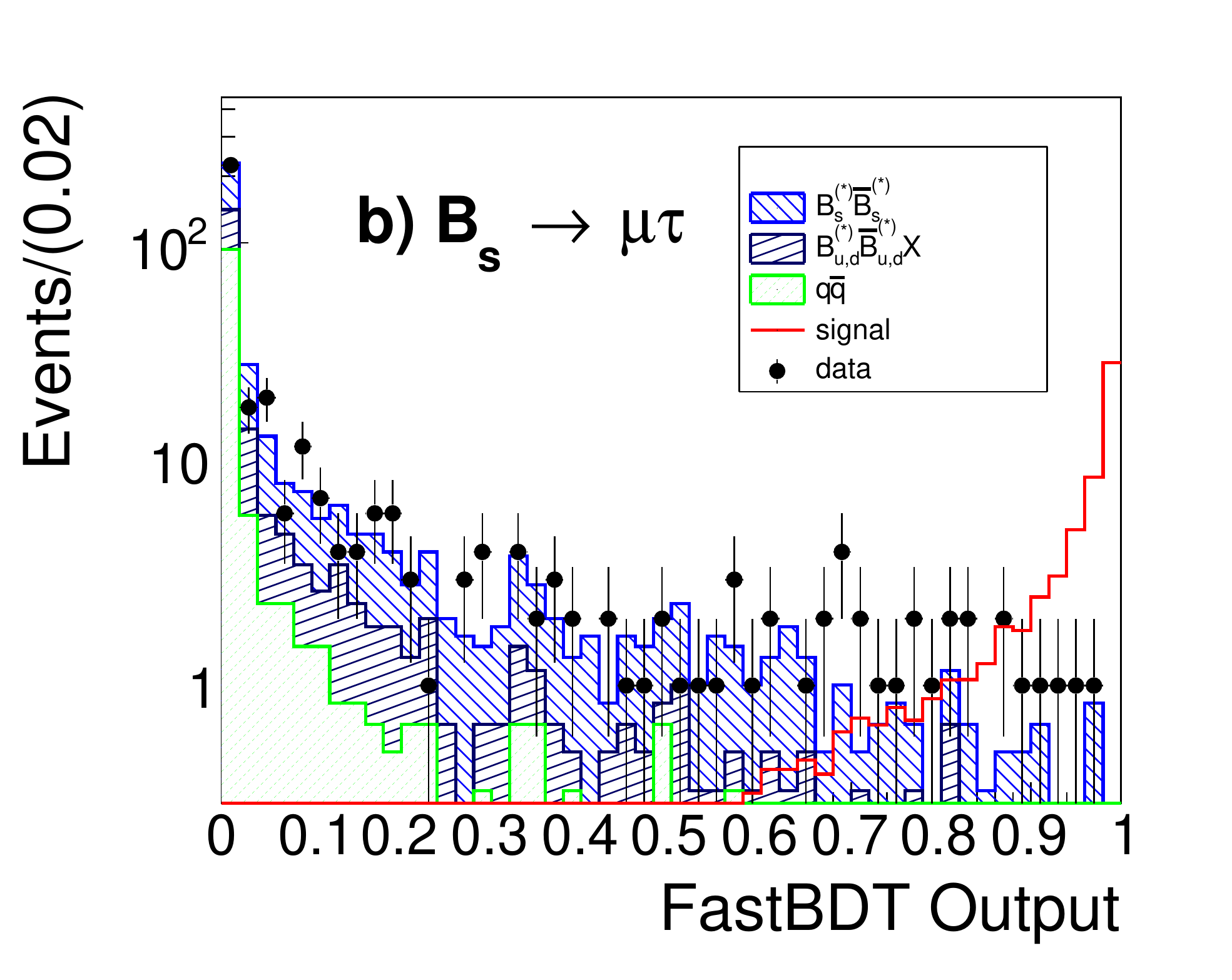}

\end{subfigure}

\end{multicols}
 \caption{The $\mathcal{O}_{\text{FastBDT}}$ distribution of {{signal MC, generic MC}} and data in $B_s \rightarrow e^-\tau^+$ (a) and $B_s \rightarrow \mu^-\tau^+$ (b) modes. The different 
 background components in generic MC are indicated by different colours as shown in the legend. The MC samples are normalized with respect to the data luminosity. The signal components correspond to $\mathcal{B} = 1 \times 10^{-2}$. The distributions are shown on a logarithmic scale.}
\label{fig: fbdtoutput}

\end{figure}

\begin{figure}[tbh]
\centering
\begin{multicols}{2}
\begin{subfigure}[t]{\linewidth}{}
\centering
\includegraphics[width= 1.1\linewidth]{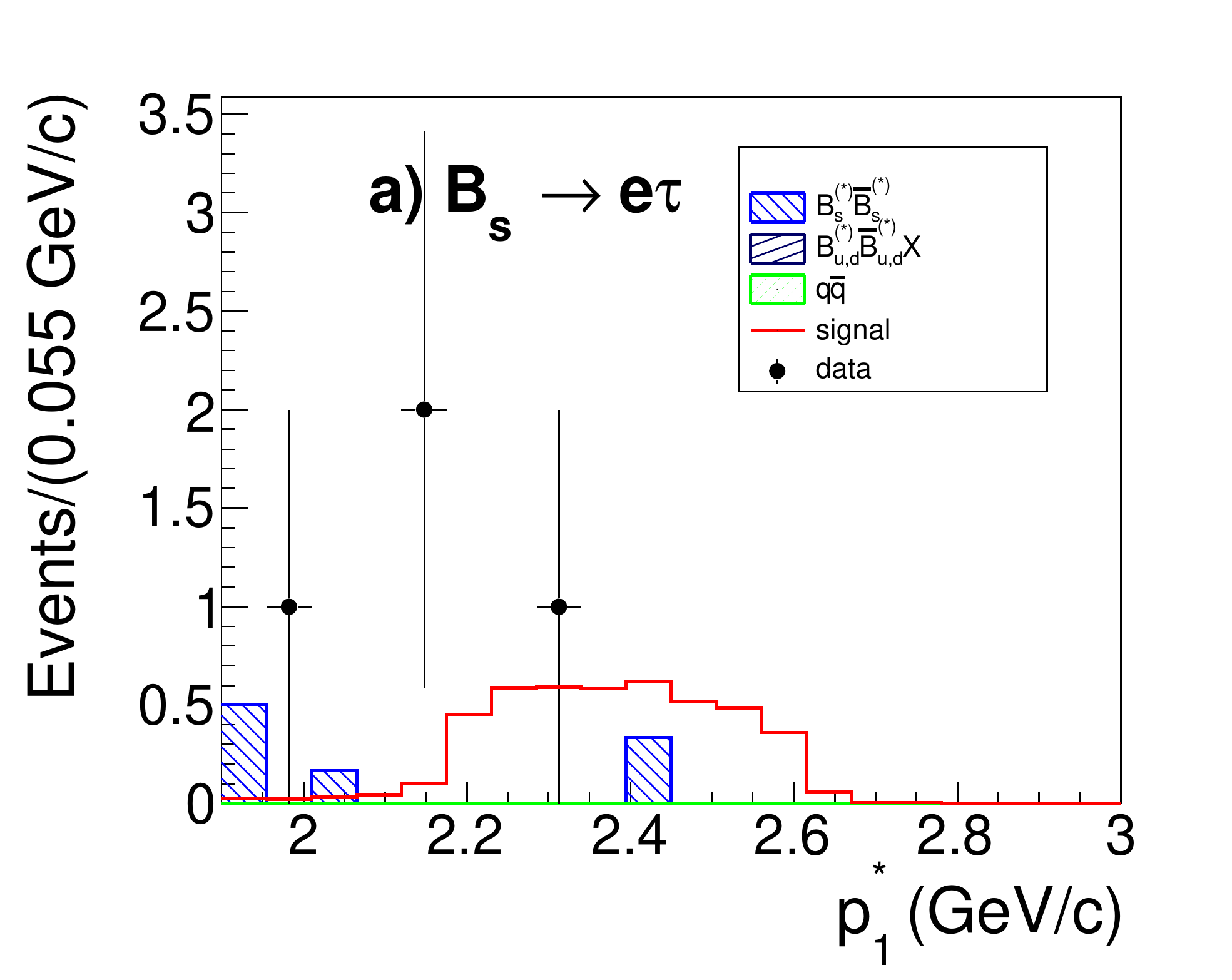}
\end{subfigure}

\begin{subfigure}[t]{\linewidth}{}
\centering
\includegraphics[width= 1.1\linewidth]{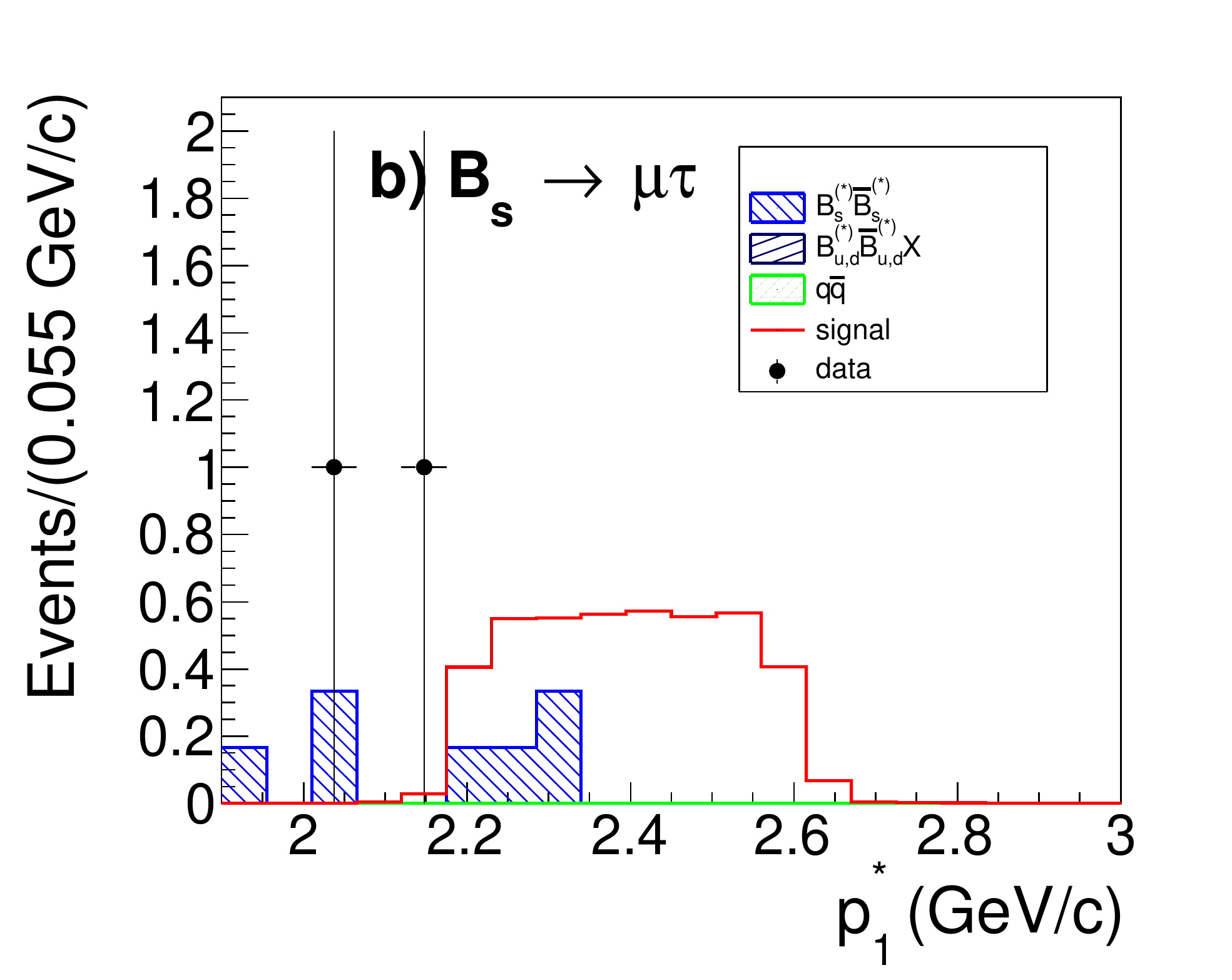}
\end{subfigure}

\end{multicols}
\caption{The $p_{1}^*$ distribution of signal MC, generic MC and data in (a) $B_s \rightarrow e^-\tau^+$ and (b) $B_s \rightarrow \mu^-\tau^+$ modes. The different 
 background components in generic MC are indicated by different colours as shown in the legend. The MC samples are normalized with respect to the data luminosity. The signal component correspond to $\mathcal{B} = 1 \times 10^{-3}$.}
\label{fig:p1}

\end{figure}

These variables do not have a significant correlation with the signal extraction variable $\pOne$. The FastBDT classifier output $\OFBDT$ ranges from zero,
where background events peak, to one, where signal events peak. For each signal mode, we choose selection criteria on $\OFBDT$
that optimize a figure-of-merit (FOM)~\cite{Punzi:2003bu}. The FOM is defined as  $\epsilon_{\text{sig}}/[(a/2)+\sqrt{N_B} ]$
, where $a=3$, $\epsilon_{\text{sig}}$ is the reconstruction efficiency of signal events as determined from MC simulation, and $N_B$ is the number of background events expected within the signal region of $\pOne \in [2.1, 2.7]~\GeV/c$. Figure~\ref{fig: fbdtoutput} shows the FastBDT output {distributions}. 
Based on FOM studies, we require $\OFBDT > 0.90$ for $B_s \rightarrow e^-\tau^+$ and $\OFBDT > 0.94$ for $B_s \rightarrow \mu^-\tau^+$ modes. These criteria reject 98$\%$ of the background events with 40$\%$ signal loss. After applying all selection criteria, 8-9$\%$ of events have
multiple signal candidates. For these events, the candidate with the highest
FastBDT output is retained. This criterion is found to select the correct signal candidate $91\%$ of the time for both decay modes. The reconstruction efficiencies from the signal simulations are
$0.032\%$ and $0.031\%$ for 
$B_s \rightarrow e^-\tau^+$ and $B_s \rightarrow \mu^-\tau^+$, respectively. Figure~\ref{fig:p1} shows the $\pOne$ distribution after applying the selection on $\OFBDT$. We observe three events for $B_s \to e^-\tau^+$ and one event for $B_s \to \mu^-\tau^+$ in the signal region.

\section{Systematic Uncertainties}
A summary of systematic uncertainties is shown in Table~\ref{tab: systematics}. 
In order to estimate the systematic uncertainty of the $\OFBDT$ selection, we use
$711~\fbi$ data sample taken at the $\Upsilon(4S)$ and reconstruct $B{}^- \rightarrow D{}^0 \pi^-$ decays, tagging the other side $B^+$ by $B{}^+ \rightarrow {\overline{D}{}^0} \ell^+ \nu$.
Here, the signal side $D{}^0$ is reconstructed in the mode $K^-\pi^+$, while the tag side $\overline{D}{}^0$
is reconstructed in  
the three decay modes $\overline{D}{}^0 \rightarrow K^+ \pi^-, \KS \pi^+ \pi^-$,  and $K^+ \pi^- \pi^+ \pi^-$.
In this study, $\pi^-$ from $B^-$ is treated as $\ell_1$ in the
signal mode, $K^-$ from $D{}^0$ as $\ell_2$, and
$\pi^+$ from $D{}^0$ is neglected. With these changes, the topology of
these events becomes similar 
to our signal mode,
and we use the same MVA as for the signal without retraining.
For this control sample study, we apply $M_{D^0} \in [1.85,1.88]~\GeV/c^2$ for both the $D^0$ in the signal side
and {$\overline{D}{}^0$} in the {tag side},
$|\DE| < 0.05~\GeV$ and $M_{\text{bc}} > 5.2~\GeV/c^2$.
Here, $\Mbc$ and $\DE$ are defined by $\Mbc = ( \sqrt{E^{*2}_{\mathrm{beam}}-|\vec{p}^{\,*}_B|^2c^2} )/c^2$ and $\DE = \EB - \Ebeam$,
where $\EB$ and $\vec{p}^{\,*}_B$ are the energy and momentum of the reconstructed $B$ meson in the c.m.\ frame. We extract the signals from a fit to $\Mbc$ with and without the $\OFBDT$ selection.
The efficiencies for $\OFBDT > 0.90$ [0.94] that are used in $B_s \to e^-\tau^+$ [$B_s \to \mu^-\tau^+$] are calculated to be $(69.3 \pm 1.7)\%$ $[(64.7 \pm 1.7)\%]$ for MC
and  $(69.9 \pm 1.6)\%$ $[(65.6 \pm 1.7)\%]$ for data. The uncertainty in the ratio of the data and MC efficiencies is assigned as the systematic uncertainty; these values are $3.3\%$ for $B_s \to e^-\tau^+$ and $3.7\%$ for $B_s \to \mu^-\tau^+$.

The semileptonic branching fraction of $B_s$ is poorly known, so
we estimate the systematic uncertainty of tagging 
from the data, using a control sample of $\overline{B}{}^0_s \rightarrow D_s^+ (X) \ell^- {\overline{\nu}_{\ell}}$,
i.e.\ the $B_s \rightarrow \ell^- \tau^+$ mode is replaced by $\Bszero \rightarrow D_s^- \ell^+ \nu_{\ell}$. 
In this control sample study, the signal side $\Bszero \rightarrow D_s^- \ell^+ \nu_{\ell}$ is 
reconstructed using three 
$D_s$ decay modes $D_s^- \rightarrow \phi(\to K^+K^-) \pi^-, \KS K^-$, and $K^{*0}(\to K^+\pi^-)K^-$.
The tag-side $\Bszerobar \rightarrow D_s^+ (X) \ell^- {\overline{\nu}_{\ell}}$ 
is reconstructed in the same way as for the $B_s \rightarrow \ell^- \tau^+$ analysis.
We require the mass of the tag-side $D_s$ meson candidate to be $1.96 < M_{D_s} <1.98~\GeV/c^2$. 
We also require the momentum of the tag-side lepton to be greater than 1.0 GeV/$c$ and $\OFBDT$ 
to be greater than 0.2. 
If there are multiple combinations in one event, the one with the highest FastBDT output is retained.
We extract the signal by performing a one-dimensional unbinned fit to $M_{D_s}$ on the signal side.
We find the signal yields $34.3 \pm 6.7$ and $37.0 \pm 6.8$ for MC and data events, respectively,
which are consistent within the uncertainty. 
These yields are approximately proportional to
the square of the tagging efficiency including 
the branching fraction of semi-leptonic $B_s$ decay to $D_s$,
so we take half the uncertainty on the yields to be the systematic uncertainty from the tag side reconstruction.
Taking into account additional contributions due to different $D_s$ reconstruction and FastBDT selection in
this control sample study, we assign $15.0~\%$ as the systematic uncertainty from the tag side reconstruction.
This uncertainty includes the contribution of the uncertainty on the branching fraction
of the semi-leptonic $B_s$ decay to $D_s$ as well as the effect of the reconstruction and selection on $D_s$ and $\ell$. 

\begin{table}[tbh]
\begin{center}
\caption{Estimated fractional systematic errors ($\%$)}
\label{tab: systematics}
\begin{tabular}{|ccc|}
\hline
Source& $B_s \rightarrow e^-\tau^+$& $B_s \rightarrow \mu^-\tau^+$\\
\hline
$\Bszerobar \rightarrow D_s^+ \ell^- \overline{\nu}_\ell$ tag &15.0 &15.0\\
FastBDT selection&3.3& 3.7\\
Lepton ID&4.3 &3.5 \\
Tracking&0.7&0.7\\
$\tau \rightarrow \ell \nu_\tau \overline{\nu}_{\ell}$ branching fraction & 0.2 & 0.2\\
 Number of $\Bs$&16.1&16.1\\
\hline
Total& 22.7& 22.6\\
\hline

\end{tabular}
\end{center}
\end{table}

Other systematic uncertainties arise from the signal-side leptons $\ell_1$ and $\ell_2$. 
The systematic uncertainty due to charged track reconstruction is estimated to be $0.35\%$ per track by using the partially reconstructed $D^{*-} \rightarrow \overline{D}{}^0\pi^-, \overline{D}{}^0\rightarrow \pi^-\pi^+\KS$ and $\KS \rightarrow \pi^+\pi^-$ events~\cite{PhysRevD.89.072009}.
The systematic uncertainties due to lepton identification are $4.3\%$ and 3.5$\%$ 
for $B_s \rightarrow e^-\tau^+$ and $B_s \rightarrow \mu^-\tau^+$ decay modes, respectively. 
The systematic uncertainties due to the $\tau$ decay branching fractions are $0.2\%$~\cite{Workman:2022ynf}.
In addition, the systematic uncertainty due to $\Bs$ meson counting is estimated as $16.1\%$. 
The total systematic uncertainty is taken as the sum in quadrature of all individual contributions.

\section{Results and Summary}
In the signal region, we find three events for $B_s \to e^-\tau^+$ and one event for $B_s \to \mu^-\tau^+$, as shown in Figure~\ref{fig:p1}.
The expected number of background events in the signal region, $\Nbkg$, is estimated from the number of events in the sideband,
scaled by the ratio of events in the signal region and sideband without the $\OFBDT$ selection as determined from MC simulation. 
Here, the sidebands are defined as $\pOne \in [1.9,2.1]$$~\GeV/c$
and $\pOne \in [2.7,3.0]~\GeV/c$.
We find $\Nbkg = 0.68 \pm 0.69$ for $B_s \to e^-\tau^+$ and $\Nbkg = 0.77 \pm 0.78$ for $B_s \to \mu^-\tau^+$.
The number of observed events in the electron mode is larger but not inconsistent with the expected number, and the probability of obtaining three or more events
with $\Nbkg = 0.68 \pm 0.69$ is $7.3\%$. The $\pOne$ distribution of the three events in the electron mode is different from the expectation for the signal.
Thus we calculate 
upper limits on the branching fractions.

To calculate this limit, we use the POLE program~\cite{PhysRevD.67.012002, pole2} with the relation 
$\mathcal{B} = (\Nobs - \Nbkg)/(\NBs \times \effsig)$, where $\Nobs$ is the number of the observed events, $\NBs$ is the number of $B_s$ mesons in the data $(16.6 \pm 2.7) \times 10^6$, and $\effsig$ is the signal efficiency including the branching fraction of $\tau$. The uncertainties on $\effsig$ and $\NBs$ listed in Section 4, together
with the uncertainty of $\Nbkg$, are taken into account in the upper limit estimation~\cite{PhysRevD.67.012002}. 
Since the uncertainty in $f_s$ is significant, 
we report the upper limit not only on the branching fraction
but also on $f_s \times \mathcal{B}(B_s \rightarrow \ell^- \tau^+)$. Table~\ref{tab:result} summarizes the results, including the upper limit.

\begin{table}[h!]
\begin{center}
\caption{Efficiency ($\epsilon$), expected background events ($N_{\mathrm{bkg}}^{\mathrm{exp}}$), observed events ($N_{\mathrm{obs}}$) and the 90$\%$ CL upper limits on $\mathcal{B}$ and $f_s \times \mathcal{B}$}
\label{tab:result}
\begin{tabular}{|cccccc|}
\hline
&$\epsilon$ ($\%$) & $N_{\mathrm{bkg}}^{\mathrm{exp}}$& $N_{\mathrm{obs}}$ & $\mathcal{B}$ & $f_s \times \mathcal{B}$   \\
& & &  &  ($\times 10^{-4}$) &  ($\times 10^{-4}$) \\
\hline
$B_s \rightarrow e^-\tau^+$ & $0.031 \pm 0.007$ & $0.68 \pm 0.69$ & $3$ & $<14$& $<5.5$  \\
$B_s \rightarrow \mu^-\tau^+$ & $0.030 \pm 0.007$ & $0.77 \pm 0.78$ & $1$ & $<7.3$ & $<2.9$ \\
\hline
\end{tabular} 
\end{center} 
\end{table}
To summarize, we have searched for the decays $\Bszero \rightarrow \ell^{\mp} \tau^{\pm}$ using the Belle data sample of 121 fb$^{-1}$ collected at the $\Upsilon(5S)$ resonance.
From the observed signal yields, we set upper limits 
\begin{eqnarray*}
\mathcal{B}(\Bszero \rightarrow e^\mp\tau^\pm) \displaystyle < 14 \times 10^{-4} \\
\mathcal{B}(\Bszero \rightarrow \mu^\mp\tau^\pm) \displaystyle < 7.3 \times 10^{-4}
\end{eqnarray*}
at $90\%$ confidence level. 
Our limit on the $\Bszero \rightarrow e^\mp\tau^\pm$ decay rate is the first such limit reported.
The sensitivity to these modes can
be improved in the future with the Belle II experiment,
which could collect a much larger data sample at the $\Upsilon(5S)$ resonance, and apply 
enhanced analysis techniques such as full
reconstruction of the tag $\overline{B}{}^0_s$~\cite{Keck:2018lcd}.

\appendix

\acknowledgments
This work, based on data collected using the Belle detector, which was
operated until June 2010, was supported by 
the Ministry of Education, Culture, Sports, Science, and
Technology (MEXT) of Japan, the Japan Society for the 
Promotion of Science (JSPS), and the Tau-Lepton Physics 
Research Center of Nagoya University; 
the Australian Research Council including grants
DP180102629, 
DP170102389, 
DP170102204, 
DE220100462, 
DP150103061, 
FT130100303; 
Austrian Federal Ministry of Education, Science and Research (FWF) and
FWF Austrian Science Fund No.~P~31361-N36;
the National Natural Science Foundation of China under Contracts
No.~11675166,  
No.~11705209;  
No.~11975076;  
No.~12135005;  
No.~12175041;  
No.~12161141008; 
Key Research Program of Frontier Sciences, Chinese Academy of Sciences (CAS), Grant No.~QYZDJ-SSW-SLH011; 
the Ministry of Education, Youth and Sports of the Czech
Republic under Contract No.~LTT17020;
the Czech Science Foundation Grant No. 22-18469S;
Horizon 2020 ERC Advanced Grant No.~884719 and ERC Starting Grant No.~947006 ``InterLeptons'' (European Union);
the Carl Zeiss Foundation, the Deutsche Forschungsgemeinschaft, the
Excellence Cluster Universe, and the VolkswagenStiftung;
the Department of Atomic Energy (Project Identification No. RTI 4002) and the Department of Science and Technology of India; 
the Istituto Nazionale di Fisica Nucleare of Italy; 
National Research Foundation (NRF) of Korea Grant
Nos.~2016R1\-D1A1B\-02012900, 2018R1\-A2B\-3003643,
2018R1\-A6A1A\-06024970, RS\-2022\-00197659,
2019R1\-I1A3A\-01058933, 2021R1\-A6A1A\-03043957,
2021R1\-F1A\-1060423, 2021R1\-F1A\-1064008, 2022R1\-A2C\-1003993;
Radiation Science Research Institute, Foreign Large-size Research Facility Application Supporting project, the Global Science Experimental Data Hub Center of the Korea Institute of Science and Technology Information and KREONET/GLORIAD;
the Polish Ministry of Science and Higher Education and 
the National Science Center;
the Ministry of Science and Higher Education of the Russian Federation, Agreement 14.W03.31.0026, 
and the HSE University Basic Research Program, Moscow; 
University of Tabuk research grants
S-1440-0321, S-0256-1438, and S-0280-1439 (Saudi Arabia);
the Slovenian Research Agency Grant Nos. J1-9124 and P1-0135;
Ikerbasque, Basque Foundation for Science, Spain;
the Swiss National Science Foundation; 
the Ministry of Education and the Ministry of Science and Technology of Taiwan;
and the United States Department of Energy and the National Science Foundation.
These acknowledgements are not to be interpreted as an endorsement of any
statement made by any of our institutes, funding agencies, governments, or
their representatives.
We thank the KEKB group for the excellent operation of the
accelerator; the KEK cryogenics group for the efficient
operation of the solenoid; and the KEK computer group and the Pacific Northwest National
Laboratory (PNNL) Environmental Molecular Sciences Laboratory (EMSL)
computing group for strong computing support; and the National
Institute of Informatics, and Science Information NETwork 6 (SINET6) for
valuable network support.
\textcolor{black}{S.N.\ is supported by JSPS KAKENHI grant JP17K05474.}

\bibliographystyle{jhep}
\bibliography{refs}{}

\end{document}